\documentclass[numbers,sort&compress,final]{elsarticle}
\usepackage[utf8]{inputenc}
\usepackage{amsmath}
\usepackage{float}

\begin{document}
\begin{frontmatter}

\title{\textbf{Renormalization group: new relations between the parameters of the Standard Model}}

\author{S.~Rebeca Juárez W.}
\ead{rebeca@esfm.ipn.mx}
\address{Departamento de Física, Escuela Superior de   Física y Matemáticas, Instituto Politécnico Nacional,  U.P. ``Adolfo López Mateos''. C.P.~07738 Ciudad de México, Mexico}

\author{Piotr Kielanowski}
\ead{kiel@fis.cinvestav.mx}
\address{Departamento de Física, Centro de Investigación y de Estudios Avanzados,
C.P. 07000 Ciudad de México, Mexico}

\author{Gerardo Mora}
\ead{gerardo.mora@ujat.mx}
\address{División Académica de Ciencias Básicas, Universidad ``Juárez'' Autónoma de Tabasco, Mexico}

\author{Arno Bohm}
\ead{bohm@physics.utexas.edu}
\address{Department of Physics, University of Texas at Austin}

\begin{abstract}
We analyze the renormalization group equations for the Standard Model at the one and two loops levels. At one loop level we find an exact constant of evolution built from the product of the quark masses and the gauge couplings $g_{1}$ and $g_{3}$ of the $U(1)$ and $SU(3)$ groups. For leptons at one loop level we find that the ratio of the charged lepton mass and the power of $g_{1}$ varies $\simeq 4\times 10^{-5}$ in the whole energy range. At the two loop level we have found two relations between the quark masses and the gauge couplings  that vary $\simeq 4\%$ and $\simeq 1\%$, respectively. For leptons at the two loop level we have derived a relation between the charged lepton mass and the gauge couplings~$g_{1}$ and~$g_{2}$ that varies $\simeq 0.1\%$. This analysis significantly simplifies the picture of the renormalization group evolution of the Standard Model and establishes new important relations between its parameters.
\end{abstract}

\end{frontmatter}

\section{Introduction}\label{sec:1}

In particle physics the renormalization group is used for the study of the asymptotic properties of the theory~\cite{ref:1,ref:2}. The renormalization group equations (RGE) for the Standard Model~\cite{ref:3,ref:4,ref:5,ref:6,ref:7,ref:8,ref:9,ref:10,ref:11,ref:12,ref:13,ref:14} is a set of coupled nonlinear differential equations, derived perturbatively, for the parameters of the theory (couplings and masses). The full set of RGE for the Standard Model is known up to two loops~\cite{ref:15} and some partial results are known at higher orders (see~\cite{ref:16} and references therein). There are no known exact solutions of the full set  of RGE for the Standard Model and only some partial results were obtained. At one loop, equations for the gauge couplings decouple and are solved exactly, but at two loops this is not the case.  Another approach, is to use the hierarchy of the parameters of the Standard model, keeping only certain powers of the quark and lepton masses and of $\lambda_{\text{CKM}}\approx0.21$ of the Cabibbo-Kobayashi-Maskawa (CKM) matrix~\cite{ref:17}. In such a way one obtains the exact solutions of the approximate equations~\cite{ref:18}. The most precise analysis of the renormalization group evolution of the Standard Model is done by numerical methods, which give very precise predictions for the evolution of the couplings, masses and CKM matrix parameters.

The aim of this paper paper is to find relations between the parameters of the Standard Model that remain constant (or are slowly varying) during the renormalization group evolution. We start with one loop equation and find an \textit{exact} constant for the quark masses and gauge couplings. Next we consider a lepton sector and find that with great accuracy ($\sim 4\times 10^{-5}$) the charged lepton masses flow proportionally to the $g_{1}^{-18/41}$. For the two loop case we find two generalizations of the one loop constant for quarks and one for leptons.

The study of the renormalization group invariants in the Standard Model and its extensions has been done before, see e.g.,~Ref~\cite{ref:22} (and references therein), where such invariants were studied for the minimal supersymmetric extension of the Standard Model. The invariants in the lepton sector were analyzed in~\cite{ref:21}. In Ref.~\cite{ref:23} an approximation of two flavors was used to simplify the problem of the analysis of complicated non linear equations. Recently, such invariants were analyzed within the powerful scheme of the flavor invariants in the minimal flavor violating extension of the Standard Model~\mbox{\cite{ref:19,ref:20}}. One should notice that all these attempts have been limited to the one loop renormalization group equations. Our approach is  different: we directly analyze the structure of the renormalization group equations, first at the one loop level and then at two loops. The invariant relations are the result of such analysis in the quark and lepton sectors.

In Sec.~\ref{sec:2} we briefly recall the renormalization group equations for the Standard Model. Next in Sec.~\ref{sec:3} we derive and discuss an \textit{exact} constant of evolution for the quark masses and gauge couplings. In Sec.~\ref{sec:4} we consider the two loop case for the quark section and derive two expressions with very slow flow. In Sec.~\ref{sec:5} we analyze slowly varying expressions in the lepton sector for one and two loop equations. We discuss our results in Sec.~\ref{sec:6}. We also include an Appendix, where we derive an equation needed in our analysis.

\section{General Considerations}\label{sec:2}

The Standard Model has the following set of parameters
$g_{1}$, $g_{2}$, $g_{3}$ -- gauge couplings,
$y_{u}$, $y_{d}$, $y_{l}$ -- Yukawa couplings of the up and down quarks and of the leptons, $\lambda$, $m$ -- Higgs quartic coupling and mass parameter. Frequently the Higgs field vacuum expectation value $v$ is used instead of $m$. All these parameters are functions of the renormalization point energy and fulfill the renormalization group equations, which have the following generic form for a parameter~$g$
\begin{equation}\label{eq:1}
\dfrac{dg}{dt}=\beta_{g}=\dfrac{1}{(4\pi)^{2}}\beta_{g}^{(1)} +\dfrac{1}{(4\pi)^{4}}\beta_{g}^{(2)}+\cdots
\end{equation}
Here $t=\ln\mu$ and $\mu$ is the renormalization point energy. The $\beta_{g}$ is the $\beta$-function, which describes the evolution of the parameter~$g$. The $\beta_{g}$~function depends on all the parameters of the theory, but does not depend explicitly on the renormalization point energy~$t$. The $\beta$-functions are calculated perturbatively and $\beta_{g}^{(i)}$ is its contribution at the $i$-loop level.

The $\beta$-functions of the gauge and Yukawa couplings and of the Higgs field vacuum expectation value~$v$ have the property that they can be factorized in the following way
\begin{equation}\label{eq:2}
\beta_{g_{i}}^{(i)}=g_{i}\tilde{\beta}_{g_{i}}^{(i)},\quad
\beta_{y_{u,d,l}}^{(i)}=y_{u,d,l}\tilde{\beta}_{y_{u,d,l}}^{(i)},\quad
\beta_{v}^{(i)}=v\tilde{\beta}_{v}^{(i)}.
\end{equation}
This means that the renormalization group equation for the Standard Model have the following form
\begin{subequations}\label{eq:3}
\begin{align}\label{eq:3a}
&\dfrac{d\ln g_{i}}{d t}=
\dfrac{1}{(4\pi)^{2}}\tilde{\beta}_{g_{i}}^{(1)}
+\dfrac{1}{(4\pi)^{4}}\tilde{\beta}_{g_{i}}^{(2)}+\cdots,\\
\label{eq:3b}&\dfrac{dy_{u,d,l}}{d t}= y_{u,d,l}\bigg(\dfrac{1}{(4\pi)^{2}}\tilde{\beta}_{u,d,l}^{(1)}
+\dfrac{1}{(4\pi)^{4}}\tilde{\beta}_{u,d,l}^{(2)}+\cdots\bigg),\\
\label{eq:3c}&\dfrac{d\ln v}{d t}=
\dfrac{1}{(4\pi)^{2}}\tilde{\beta}_{v}^{(1)}
+\dfrac{1}{(4\pi)^{4}}\tilde{\beta}_{v}^{(2)}+\cdots.
\end{align}
\end{subequations}

Yukawa couplings $y_{u,d,l}$ are complex matrices $3\times 3$, so Eq.~\eqref{eq:3b} is a matrix differential equation. Yukawa couplings  $y_{u,d}$ couple to the left- and right-handed quarks. Let us observe that the diagonalization of the matrix $y_{u,d}$ by a biunitary transformation requires also the knowledge of the right diagonalizing matrix, which is not related to any observable of the Standard Model. For this reason it is more convenient to use the matrices $H_{u,d}=y^{\dagger}_{u,d}y^{\phantom{\dagger}}_{u,d}$ which are hermitian and are diagonalized by the left diagonalizing unitary matrices. which are related to the Cabibbo-Kobayashi-Maskawa matrix. It follows from Eq.~\eqref{eq:3} that the matrices $H_{u,d}=y^{\dagger}_{u,d}y^{\phantom{\dagger}}_{u,d}$ fulfill the following differential equation
\begin{multline}\label{eq:4}
\dfrac{d H_{u,d}}{d t} 
=H_{u,d}\bigg(\dfrac{1}{(4\pi)^{2}}
{{\tilde{\beta}^{(1)}_{u,d}}}\phantom{)^{\dagger}} +\dfrac{1}{(4\pi)^{4}}{{\tilde{\beta}^{(2)}_{u,d}}} \phantom{)^{\dagger}} +\cdots
\bigg)\\
+\bigg(\dfrac{1}{(4\pi)^{2}}
{{\tilde{\beta}^{(1)}_{u,d}}}\phantom{)^{\dagger}} +\dfrac{1}{(4\pi)^{4}}{{\tilde{\beta}^{(2)}_{u,d}}} \phantom{)^{\dagger}} +\cdots
\bigg)^{\dagger}H_{u,d}\\
=
\dfrac{1}{(4\pi)^{2}}\Big(H_{u,d} {{\tilde{\beta}^{(1)}_{u,d}}}\phantom{)}^{\phantom{\dagger}} + ({{\tilde{\beta}^{(1)}_{u,d}}})^{\dagger} H_{u,d}\Big)
+\dfrac{1}{(4\pi)^{4}}\Big(H_{u,d} {{\tilde{\beta}^{(2)}_{u,d}}}\phantom{)}^{\phantom{\dagger}} + ({{\tilde{\beta}^{(2)}_{u,d}}})^{\dagger} H_{u,d}\Big)+\cdots,
\end{multline}
and the symbol $\dagger$ means here the hermitian conjugate matrix.

The one loop approximation consists in keeping only the terms $\tilde{\beta}^{(1)}$ on the right hand side of Eqs.~\eqref{eq:3} and~\eqref{eq:4}, the two loop approximation consists in keeping the terms $\tilde{\beta}^{(1)}$ and $\tilde{\beta}^{(2)}$ on the right hand side of Eqs.~\eqref{eq:3} and~\eqref{eq:4}, etc.

\section{A constant for the one loop evolution}\label{sec:3}
\subsection{Derivation}\label{sec:3.1}
The explicit form of the one loop $\beta^{(1)}$ functions for the gauge and Yukawa couplings and for the Higgs field vacuum expectation value~$v$ is the following
\begin{subequations}\label{eq:5}
\begin{align}&\label{eq:5a}
\tilde{\beta}^{(1)}_{g_{i}}=-b_{i}g_{i}^{2},\quad \{b_{1},b_{2},b_{3}\} = \{-\dfrac{41}{10},\dfrac{19}{6},7\},\\
&\label{eq:5b}
\tilde{\beta}^{(1)}_{y_{u}}= \dfrac{3}{2}(y^{\dagger}_{u}y^{\phantom{\dagger}}_{u} -y^{\dagger}_{d}y^{\phantom{\dagger}}_{d}) +Y_{2}(S) -\left(\dfrac{17}{20}g_{1}^{2}+\dfrac{9}{4}g_{2}^{2} +8g_{3}^{2}\right),
\\
&\label{eq:5c}
\tilde{\beta}^{(1)}_{y_{d}}= \dfrac{3}{2}(y^{\dagger}_{d}y^{\phantom{\dagger}}_{d} -y^{\dagger}_{u}y^{\phantom{\dagger}}_{u}) +Y_{2}(S) -\left(\dfrac{1}{4}g_{1}^{2}+\dfrac{9}{4}g_{2}^{2} +8g_{3}^{2}\right),
\\
&\label{eq:5cl}
\tilde{\beta}^{(1)}_{y_{l}}=
\dfrac{3}{2}y^{\dagger}_{l}y^{\phantom{\dagger}}_{l}+Y_{2}(S) -\dfrac{9}{4}(g_{1}^{2}+g_{2}^{2}),\\
&\label{eq:5d}
\tilde{\beta}_{v}^{(1)}=\dfrac{9}{4}\left(\dfrac{1}{5}g_{1}^{2}+g_{2}^{2}\right)-Y_{2}(S),
\\
&\nonumber
Y_{2}(S)=\operatorname{Tr}(3y^{\dagger}_{u}y^{\phantom{\dagger}}_{u} +3y^{\dagger}_{d}y^{\phantom{\dagger}}_{d} +y^{\dagger}_{l}y^{\phantom{\dagger}}_{l}).
\end{align}
\end{subequations}

Now let us analyze the RGE for the $H_{u,d}$ matrices couplings. From Eq.~\eqref{eq:4} we know that they are the first order ordinary differential matrix equations of very specific form discussed in~\ref{app:A}. From this discussion it follows that one can derive equations for the determinant of the $H_{u}$ and $H_{d}$ matrices:
\begin{subequations}\label{eq:6}
\begin{align}
&\label{eq:6a}
\dfrac{d\det H_{u}}{d t} = \dfrac{1}{(4\pi)^{2}}\det H_{u}(\operatorname{Tr}(\tilde{\beta}_{u}^{(1)}\vphantom{)^{\dagger}} +(\tilde{\beta}_{u}^{(1)})^{\dagger})\Rightarrow\nonumber\\
&\phantom{\dfrac{d\det H_{u}}{d t} = \dfrac{1}{(4\pi)^{2}}\det H_{u}}\dfrac{d\ln(\det H_{u})}{dt}
 = \dfrac{1}{(4\pi)^{2}}\operatorname{Tr}(\tilde{\beta}_{u}^{(1)}
\vphantom{)^{\dagger}} +(\tilde{\beta}_{u}^{(1)})^{\dagger}),\\
&\label{eq:6b}
\dfrac{d\det H_{d}}{d t} = \dfrac{1}{(4\pi)^{2}}\det H_{d}(\operatorname{Tr}(\tilde{\beta}_{d}^{(1)}\vphantom{)^{\dagger}} +(\tilde{\beta}_{d}^{(1)})^{\dagger})\Rightarrow\nonumber\\ &\phantom{\dfrac{d\det H_{d}}{d t} = \dfrac{1}{(4\pi)^{2}}\det H_{d}}\dfrac{d\ln(\det H_{d})}{dt}
 = \dfrac{1}{(4\pi)^{2}}\operatorname{Tr}(\tilde{\beta}_{d}^{(1)}
\vphantom{)^{\dagger}} +(\tilde{\beta}_{d}^{(1)})^{\dagger}).
\end{align}
\end{subequations}
Thus, taking into account Eqs.~\eqref{eq:3}, \eqref{eq:5} and~\eqref{eq:6} we obtain at the one loop level
\begin{equation}\label{eq:7}
\dfrac{d\ln\det H_{u}}{dt} +\dfrac{d\ln\det H_{u}}{d t} +12 \dfrac{d\ln v}{d t}= \dfrac{1}{(4\pi)^{2}}\left(-\dfrac{6}{5}g_{1}^{2}-96 g_{3}^{2}\right).
\end{equation}
Now, from Eqs.~\eqref{eq:3} and~\eqref{eq:5a} we have
\begin{equation}\label{eq:8}
\dfrac{1}{(4\pi)^{2}}g_{i}^{2}= -\dfrac{1}{b_{i}}\dfrac{d\ln g_{i}}{d t}
\end{equation}
and Eq.~\eqref{eq:7} can be rewritten as
\begin{multline}\label{eq:9}
\dfrac{d}{d t}
\left(\ln \dfrac{\det H_{u}\det H_{d}v^{12}} {g_{1}^{\frac{6}{5b_{1}}}g_{3}^{\frac{96}{b_{3}}}}\right)\\ =\dfrac{d\ln\det H_{u}}{d t} +\dfrac{d\ln\det H_{u}}{d t} +12 \dfrac{d\ln v}{d t} -\dfrac{6}{5b_{1}} \dfrac{d\ln g_{1}}{d t} -\dfrac{96}{b_{3}} \dfrac{d\ln g_{3}}{d t}\\
=\dfrac{1}{(4\pi)^{2}}\Big(\operatorname{Tr}(\tilde{\beta}_{u}^{(1)}\vphantom{)^{\dagger}} +(\tilde{\beta}_{u}^{(1)})^{\dagger} +\tilde{\beta}_{d}^{(1)}\vphantom{)^{\dagger}} +(\tilde{\beta}_{d}^{(1)})^{\dagger}+12\tilde{\beta}_{v}^{(1)} -\dfrac{6}{5b_{1}}\tilde{\beta}_{g_{1}}^{(1)} -\dfrac{96}{b_{3}}\tilde{\beta}_{g_{3}}^{(1)}\Big) =0.
\end{multline}
This means that the following function of the parameters of the Standard Model is constant upon the renormalization group evolution
\begin{equation}\label{eq:10}
\dfrac{\det H_{u}\det H_{d}v^{12}} {g_{1}^{\frac{6}{5b_{1}}}g_{3}^{\frac{96}{b_{3}}}}=\text{const.}
\end{equation}

Let us now express the constant in Eq.~\eqref{eq:10} in terms of observables. The eigenvalues of the hermitian matrices $H_{u,d} =y^{\dagger}_{u,d}y^{\phantom{\dagger}}_{u,d}$ are the squares of the eigenvalues of the Yukawa coupling matrices, corresponding to the up- and down- quarks $\{Y_{t}^{2},Y_{c}^{2},Y_{u}^{2}\}$ and $\{Y_{b}^{2},Y_{s}^{2},Y_{d}^{2}\}$. The determinants of $H_{u}$ and $H_{d}$ are thus equal
\begin{equation}\label{eq:11}
\det H_{u} = \left(Y_{t}Y_{c}Y_{u}\right)^{2}\text{ and } \det H_{d} = \left(Y_{b}Y_{s}Y_{d}\right)^{2}.
\end{equation}
The quark masses are equal
\begin{equation}\label{eq:12}
m_i=\dfrac{Y_{i}v}{\sqrt{2}}.
\end{equation}
Taking into account Eqs.~\eqref{eq:11} and~\eqref{eq:12} the evolution constant from Eq.~\eqref{eq:10} can be rewritten in terms of the quark masses and gauge couplings
\begin{equation}\label{eq:13}
64\left(\dfrac{m_{t}m_{c}m_{u}m_{b}m_{s}m_{d}} {g_{1}^{\frac{3}{5b_{1}}}g_{3}^{\frac{48}{b_{3}}}}\right)^{2} =\text{const.}
\end{equation}
Thus $K_{1}$ defined below is the one loop constant of the renormalization group evolution
\begin{equation}\label{eq:14}
K_{1}=\dfrac{m_{t}m_{c}m_{u}m_{b}m_{s}m_{d}} {g_{1}^{\frac{3}{5b_{1}}}g_{3}^{\frac{48}{b_{3}}}}= \dfrac{m_{t}m_{c}m_{u}m_{b}m_{s}m_{d}} {g_{1}^{-\frac{6}{41}}g_{3}^{\frac{48}{7}}}=\text{const.}
\end{equation}
\subsection{Numerical analysis of the one loop constant $K_{1}$}\label{sec:3.2}

\begin{figure}[h!]
\centering
\includegraphics{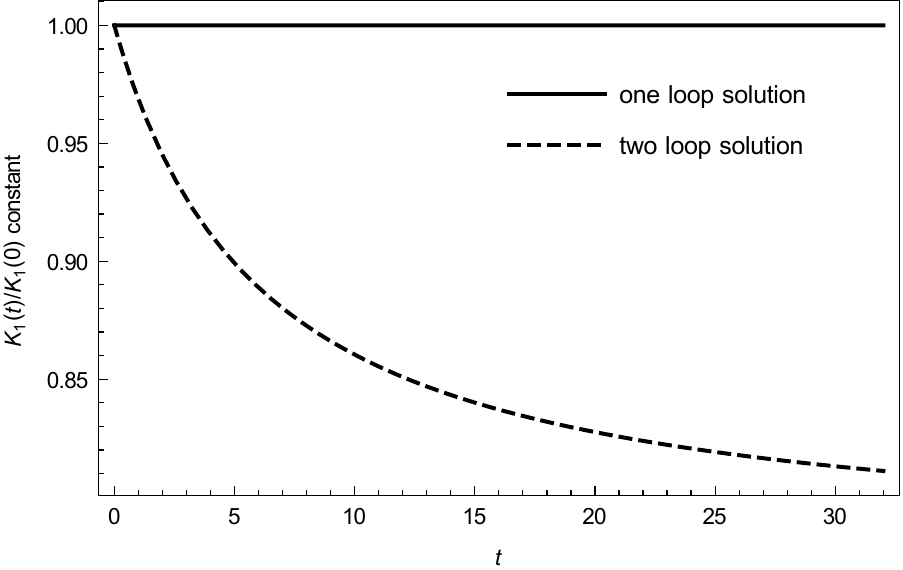}
\caption{Plot of the $K_{1}$ relation (which is constant at the one loop level) normalized to $1$ at $t=0$. The line labeled ``one loop solution'' has been drawn using the numerical solution of the renormalization group equations at the \textit{one} loop level. The line labeled ``two loop solution'' has been drawn using the numerical solution of the renormalization group equations at the \textit{two} loop level. \label{fig:1}}
\end{figure}

We will now display the evolution of the constant $K_{1}$ from Eq.~\eqref{eq:14}, using the evolution of the parameters $m_{i}$ and~$g_{i}$ obtained from the numerical solution of the one and two loop renormalization group equations. The results are shown in Fig.~\ref{fig:1}, where we draw the constant $K_{1}$ normalized to~1 at $t=0$ (by dividing it by its value at $t=0$). The one loop evolution of $K_{1}$ produces a perfect straight line with a constant value, which demonstrates that the numerical analysis is compatible with the analytical one. The $K_{1}$ relation at the two loop solution of the renormalization group equations is not constant, what mathematically is expected, because $K_{1}$ was derived from the one loop equations. However it is rather surprising that the variation is so large, more that~15\%. We will discuss it later after the analysis of the two loop equations.

\section{Discussion of the two loop evolution}\label{sec:4}

The evolution of the constant $K_{1}$, at one loop level, given in Fig.~\ref{fig:1} shows that $K_{1}$ is not constant for the two loop solution of the renormalization group equations. Unfortunately the two loop equations are too complicated to derive analytically another quantity that might be constant. However, we will analyze the situation and we will introduce some improvements.

From Eq.~\eqref{eq:9} generalized to two loops it follows that
\begin{multline}\label{eq:15}
\dfrac{d}{d t}
\left(\ln \dfrac{\det H_{u}\det H_{d}v^{12}} {g_{1}^{\frac{6}{5b_{1}}}g_{3}^{\frac{96}{b_{3}}}}\right)\\ =
\dfrac{1}{(4\pi)^{4}}\Big(\operatorname{Tr}(\tilde{\beta}_{u}^{(2)}\vphantom{)^{\dagger}} +(\tilde{\beta}_{u}^{(2)})^{\dagger} +\tilde{\beta}_{d}^{(2)}\vphantom{)^{\dagger}} +(\tilde{\beta}_{d}^{(2)})^{\dagger})+12\tilde{\beta}_{v}^{(2)} -\dfrac{6}{5b_{1}}\tilde{\beta}_{g_{1}}^{(2)} -\dfrac{96}{b_{3}}\tilde{\beta}_{g_{3}}^{(2)}\Big)\\
=\dfrac{1}{(4\pi)^{4}}\Bigg(k_{1}g_{1}^{4}+ k_{2}g_{2}^{4} +k_{3}g_{3}^{4} +k_{4}g_{1}^{2}g_{2}^{2}+ k_{5}g_{1}^{2}g_{3}^{2}+ k_{6}g_{2}^{2}g_{3}^{2}
+\dfrac{17}{2}\operatorname{Tr}(H_{u}^{2}+H_{d}^{2})\\ -5\operatorname{Tr}(H_{u}H_{d}) + \Big(\dfrac{636}{205} g_{1}^{2}+18 g_{2}^{2}+\dfrac{192}{7} g_{3}^{2}-16 \lambda- 2Y_{2}(S)\Big)\operatorname{Tr}H_{u}\\ +\Big(\dfrac{708}{205} g_{1}^{2}+18 g_{2}^{2}+\dfrac{192}{7} g_{3}^{2}-16 \lambda-2 Y_{2}(S)\Big)\operatorname{Tr}H_{d} -\dfrac{18}{41}g_{1}^{2}\operatorname{Tr}H_{l} \Bigg) \neq 0,
\end{multline}
\begin{equation*}
k_{1}=-\dfrac{62567}{8200},\; k_{2}=\dfrac{261}{8},\; k_{3}=-\dfrac{6576}{7},\; k_{4}= -\dfrac{11529}{820},\; k_{5}=\dfrac{10748}{1435},\; k_{6}=\dfrac{324}{7}.
\end{equation*}
Here $\lambda$ is the Higgs quartic coupling.
The right hand side of Eq.~\eqref{eq:15} cannot be analytically expressed as a derivative and thus it is not possible to find a two loop analogue of constant $K_{1}$. However the terms of the type $g_{i}^{4}$ and $g_{i}^{2}g_{j}^{2}$ can be expressed as derivatives using the one loop renormalization group equations
\begin{subequations}\label{eq:16}
\begin{align}\label{eq:16a}
&\dfrac{1}{(4\pi)^{4}}g_{i}^{4}= -\dfrac{1}{(4\pi)^{2}} \dfrac{1}{2b_{i}}\dfrac{d\,g_{i}^{2}}{dt} =\dfrac{d}{dt} \Bigg(\ln \exp\bigg(-\frac{1}{2b_{i}}\dfrac{1}{(4\pi)^{2}}g_{i}^{2}\bigg) \Bigg),
\\
&\dfrac{1}{(4\pi)^{4}}g_{i}^{2}g_{j}^{2}= \dfrac{1}{(4\pi)^{4}}\dfrac{g_{i0}^{2}g_{j0}^{2}} {b_{i}g_{i0}^{2}-b_{j}g_{j0}^{2}}(b_{i}g_{i}^{2}-b_{j}g_{j}^{2}) \nonumber\\
&\phantom{\dfrac{1}{(4\pi)^{4}}g_{i}^{2}g_{j}^{2}} = -\dfrac{1}{(4\pi)^{2}}
\dfrac{g_{i0}^{2}g_{j0}^{2}} {b_{i}g_{i0}^{2}-b_{j}g_{j0}^{2}}
\Bigg(\dfrac{d\ln g_{i}}{dt}-\dfrac{d\ln g_{j}}{dt}\Bigg)\nonumber\\
&\label{eq:16b}
\phantom{\dfrac{1}{(4\pi)^{4}}g_{i}^{2}g_{j}^{2}} =
\dfrac{d}{dt}\Bigg( \ln\bigg(\dfrac{g_{i}}{g_{j}}\bigg) ^{-\frac{1}{(4\pi)^{2}}
\frac{g_{i0}^{2}g_{j0}^{2}} {b_{i}g_{i0}^{2}-b_{j}g_{j0}^{2}}} \Bigg).
\end{align}
\end{subequations}
Here $g_{i0}$ is the value of the $g_{i}$ coupling at $t=0$.

\begin{figure}[h!]
\centering
\includegraphics{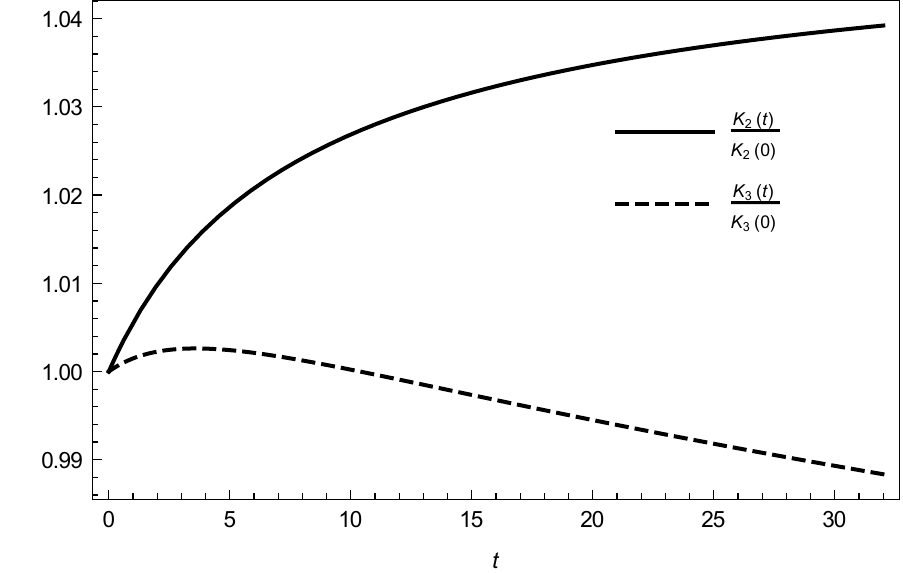}
\caption{Plot of the evolution of the two quantities~$K_{2}$ and~$K_{3}$ normalized to~1 at $t=0$. This figure should be compared with Fig.~\ref{fig:1}, where the constant $K_{1}$ for the two loop evolution shows a significant variation. $K_{2}$ changes only 4\% in the whole range of the energy and~$K_{3}$ is more stable than $K_{2}$ and it changes only 1\%. \label{fig:2}}
\end{figure}

Now, inserting Eqs.~\eqref{eq:16} into Eq.~\eqref{eq:15} and moving all derivatives to the left hand side we obtain
\begin{multline}\label{eq:17}
\dfrac{d}{d t}
\left(\ln \dfrac{\det H_{u}\det H_{d}v^{12}} {g_{1}^{\frac{6}{5b_{1}}+r_{1}}
g_{2}^{r_{2}}
g_{3}^{\frac{96}{b_{3}}+r_{3}}\exp(-\frac{1}{2(4\pi)^{2}}(\frac{k_{1}}{b_{1}}g_{1}^{2} +\frac{k_{2}}{b_{2}}g_{2}^{2}+\frac{k_{3}}{b_{3}}g_{3}^{2}))}\right)\\
=\dfrac{1}{(4\pi)^{4}}\Bigg(\dfrac{17}{2}\operatorname{Tr}(H_{u}^{2}+H_{d}^{2}) -5\operatorname{Tr}(H_{u}H_{d})\\ + \Big(\dfrac{636}{205} g_{1}^{2}+18 g_{2}^{2}+\dfrac{192}{7} g_{3}^{2}-16 \lambda- 2Y_{2}(S)\Big)\operatorname{Tr}H_{u}\\ +\Big(\dfrac{708}{205} g_{1}^{2}+18 g_{2}^{2}+\dfrac{192}{7} g_{3}^{2}-16 \lambda-2 Y_{2}(S)\Big)\operatorname{Tr}H_{d} -\dfrac{18}{41}g_{1}^{2}\operatorname{Tr}H_{l} \Bigg),
\end{multline}
\begin{align*}
r_{1}&= - \frac{1}{(4\pi)^{2}}\bigg(
\dfrac{k_{4}g_{10}^{2}g_{20}^{2}}
{b_{1}g_{10}^{2}-b_{2}g_{20}^{2}}
+ 
\dfrac{k_{5}g_{10}^{2}g_{30}^{2}}
{b_{1}g_{10}^{2}-b_{3}g_{30}^{2}}\bigg),\\
r_{2}&=\frac{1}{(4\pi)^{2}}\bigg(
\dfrac{k_{4}g_{10}^{2}g_{20}^{2}}
{b_{1}g_{10}^{2}-b_{2}g_{20}^{2}}
- 
\dfrac{k_{6}g_{20}^{2}g_{30}^{2}}
{b_{2}g_{20}^{2}-b_{3}g_{30}^{2}}\bigg),\\
r_{3}&= \frac{1}{(4\pi)^{2}}\bigg(
\dfrac{k_{5}g_{10}^{2}g_{30}^{2}}
{b_{1}g_{10}^{2}-b_{3}g_{30}^{2}}
+ 
\dfrac{k_{6}g_{20}^{2}g_{30}^{2}}
{b_{3}g_{30}^{2}-b_{3}g_{30}^{2}}\bigg).
\end{align*}
Using Eq.~\eqref{eq:17} we define
\begin{equation}\label{eq:18}
K_{2}=\dfrac{m_{t}m_{c}m_{u}m_{b}m_{s}m_{d}} {g_{1}^{\frac{3}{5b_{1}}+\frac{r_{1}}{2}}
g_{2}^{\frac{r_{2}}{2}}
g_{3}^{\frac{48}{b_{3}}+\frac{r_{3}}{2}}\exp\big(-\frac{1}{4(4\pi)^{2}}(\frac{k_{1}}{b_{1}}g_{1}^{2} +\frac{k_{2}}{b_{2}}g_{2}^{2}+\frac{k_{3}}{b_{3}}g_{3}^{2})\big)}.
\end{equation}
Let us also define the quantity $K_{3}$, which is a slight modification of $K_{2}$
\begin{equation}\label{eq:19}
K_{3}=\dfrac{m_{t}m_{c}m_{u}m_{b}m_{s}m_{d}} {g_{1}^{\frac{3}{5b_{1}}+\frac{b_{1}r_{1}}{2}}
g_{2}^{\frac{b_{2}r_{2}}{2}}
g_{3}^{\frac{96}{b_{3}}+\frac{b_{3}r_{3}}{2}}\exp\big(-\frac{1}{4(4\pi)^{2}}(\frac{k_{1}}{b_{1}}g_{1}^{2} +\frac{k_{2}}{b_{2}}g_{2}^{2}+\frac{k_{3}}{b_{3}}g_{3}^{2})\big)}.
\end{equation}
The two loop evolution of the quantities $K_{2}$ and $K_{3}$ normalized to~1 at $t=0$ is shown in Fig.~\ref{fig:2}

\section{Lepton case}\label{sec:5}

\begin{figure}[h!]
\includegraphics[width=0.47\linewidth]{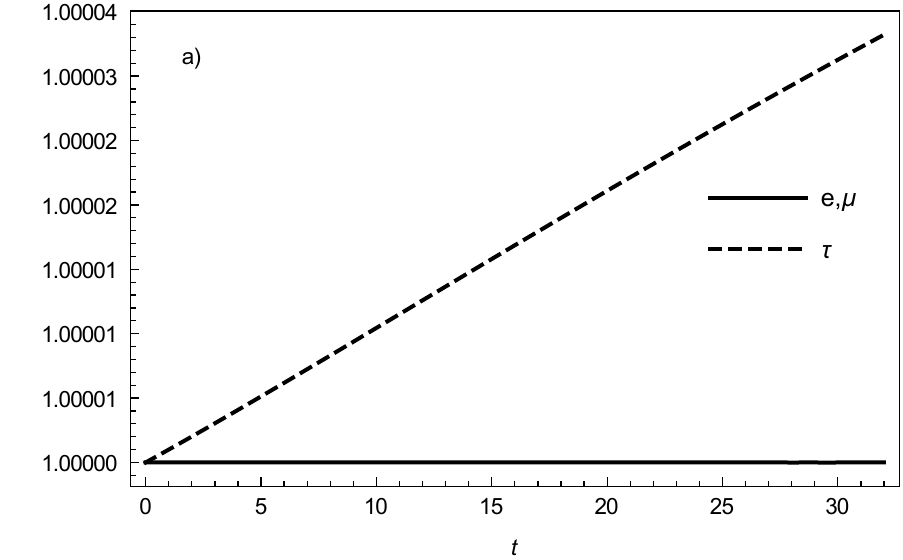}\hspace*{0.05\linewidth}
\includegraphics[width=0.47\linewidth]{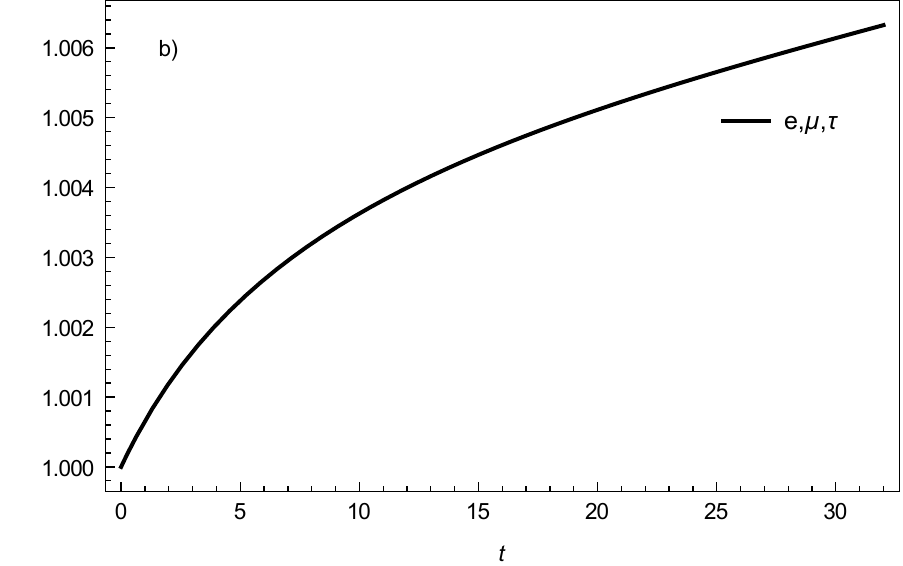}
\caption{Plot of the approximate one loop constant for leptons. In~a) we draw the evolution of $K_{4}$ using the one loop solution of the renormalization group equations and in~b) we draw the evolution of $K_{4}$ for the two loop solution.\label{fig:3}}
\end{figure}

The matrix of the Yukawa couplings for leptons is diagonal\linebreak $y_{l}=\operatorname{Diag(Y_{e},Y_{\mu},Y_{\tau})}$ and the renormalization group equations for the lepton sector decouple, so one does not have to consider the equation for the determinant. If one takes the sum of Eqs.~\eqref{eq:5cl} and~\eqref{eq:5d} and uses Eq.~\eqref{eq:8} one obtains
\begin{multline}\label{eq:20}
\dfrac{d\ln Y_{i}}{dt} +\dfrac{d\ln v}{dt} =\dfrac{1}{(4\pi)^{2}}\Big(\dfrac{3}{2}Y_{i}^{2}-\dfrac{9}{5}g_{1}^{2}\Big)\Rightarrow\\
\dfrac{d}{dt}\left(\ln\dfrac{m_{i}}{g_{1}^{\frac{9}{5b_{1}}}}\right) =\dfrac{3}{2(4\pi)^{2}}Y_{i}^{2}, \quad i=\{e,\mu,\tau\}
\end{multline}
and we define the quantity~$K_{4}$, which is approximate constant for leptons
\begin{equation}\label{eq:21}
K_{4}=\dfrac{m_{i}}{g_{1}^{\frac{9}{5b_{1}}}}= \dfrac{m_{i}}{g_{1}^{-\frac{18}{41}}}= \exp\bigg(\frac{1}{(4\pi)^{2}}
\int_{t_{0}}^{t}Y_{i}^{2}dt\bigg)\approx \text{ const.},\quad i=\{e,\mu,\tau\}.
\end{equation}
In Fig.~\ref{fig:3} we plot the evolution of $K_{4}$. In Fig.~\ref{fig:3}~a) we use the one loop evolution and one sees that for electron and muon one cannot notice any variation and for lepton $\tau$ the evolution is linear but very small. In Fig.~\ref{fig:3}~b) we plot $K_{4}$ and use the two loop solution of the renormalization group equations and  one cannot notice any difference between electron, muon and $\tau$.

\begin{figure}[H]
\centering
\includegraphics{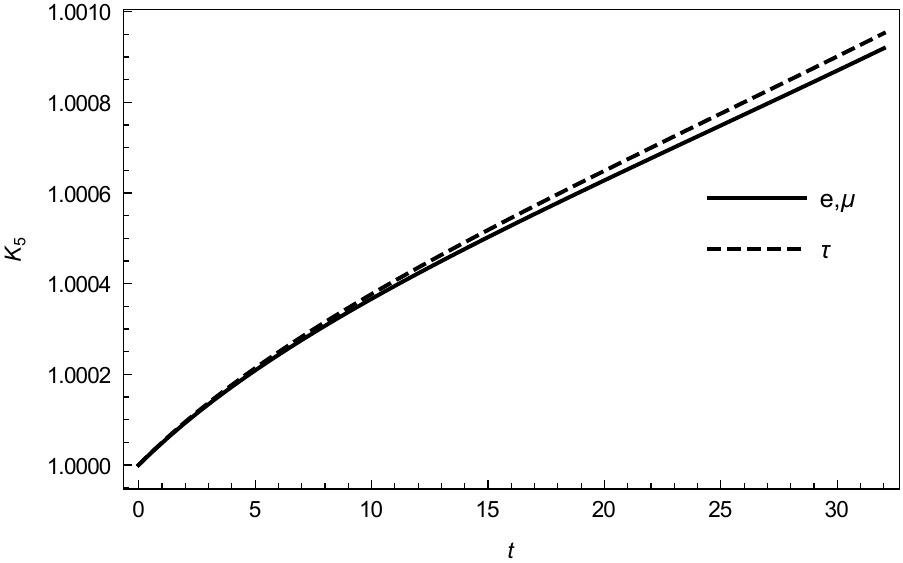}
\caption{The plot of the relation $K_{5}$ given in Eq.~\eqref{eq:23} for electron, $\mu$ and~$\tau$. There is no visible difference between the evolution of $K_{5}$ for electron and $\mu$. One can see the significant improvement in comparison with Fig.~\ref{fig:3}b. \label{fig:4}}
\end{figure}

The two loop analogue of Eq.~\eqref{eq:20} has the following form
\begin{multline}\label{eq:22}
\bigg(\dfrac{d\ln Y_{i}}{dt} +\dfrac{d\ln v}{dt}\bigg)\bigg|_{\text{{two loop}}}\\ =\dfrac{1}{(4\pi)^{4}}\Big(k_{7}g_{1}^{4} +k_{8}g_{2}^{4} +k_{9}g_{1}^{2}g_{2}^{2} +\dfrac{3}{2}Y_{i}^{4}+ \big(\dfrac{387}{80}g_{1}^{2} +\dfrac{135}{16}g_{2}^{2}-\dfrac{9}{4} Y_{2}(S)-6\lambda\big)Y_{i}^{2} \Big),
\end{multline}
\begin{equation*}
k_{7}=\dfrac{4191}{800},\; k_{8}=\dfrac{87}{32},\; k_{9}=\dfrac{81}{80}.
\end{equation*}
and the two loop approximate lepton relation $K_{5}$ is equal
\begin{multline}\label{eq:23}
K_{5}=\dfrac{m_{i}}{g_{1}^{\frac{9}{5b_{1}}+r_{4}}g_{2}^{-r_{4}}
\exp\big(-\frac{1}{2(4\pi)^{2}}(\frac{k_{7}}{b_{1}}g_{1}^{2} +\frac{k_{8}}{b_{2}}g_{2}^{2})\big)}\\
 =\exp\bigg(\dfrac{1}{(4\pi)^{2}}\int_{t_{0}}^{t} Y_{i}^{2}dt +\dfrac{1}{(4\pi)^{4}}\int_{t_{0}}^{t}\Big(\dfrac{3}{2}Y_{i}^{4}+ \big(\dfrac{387}{80}g_{1}^{2} +\dfrac{136}{16}g_{2}^{2}-\dfrac{9}{4} Y_{2}(S)-6\lambda\big)Y_{i}^{2} \Big)dt\bigg)\\
 \approx\text{const.}
, \quad i=\{e,\mu,\tau\},
\end{multline}
\begin{equation*}
r_{4}= - \frac{1}{(4\pi)^{2}}
\dfrac{k_{9}g_{10}^{2}g_{20}^{2}}
{b_{1}g_{10}^{2}-b_{2}g_{20}^{2}}.
\end{equation*}
In Fig.~\ref{fig:4} we plot the relation $K_{5}$  given in Eq.~\eqref{eq:23} for the two loop solution of the renormalization group equations. One can see that the variation of $K_{5}$ in the whole range of energy is approximately 0.1\%, so with great accuracy one can say that $K_{5}$ behaves like a constant.

\section{Discussion of the results}\label{sec:6}

We have analyzed the renormalization group equations for the Standard Model and we have derived various expressions built from the parameters of the theory that are constant or \textit{almost} constant during the evolution. The form of the expressions depends whether we use the one or two loops equations.

In case of the one loop equations the evolution constant $K_{1}$ is given in Eq.~\eqref{eq:14}. The numerator of $K_{1}$ is the product of the quark masses and the denominator is equal to the product $g_{1}^{\frac{6}{41}}g_{3}^{\frac{48}{7}}$ ($g_{1}$ and $g_{3}$ are the gauge couplings of $U(1)$ and $SU(3)$, respectively). This means that the dependence on $g_{2}$ of the product of quark masses cancels out. This cancellation is interesting, because it shows that the product of masses of all quarks depends only on electromagnetic and strong interactions.

For leptons there is no exact constant of one loop evolution, but the relation $K_{4}$, for each charged lepton, given in Eq.~\eqref{eq:21} has very small variation in the whole energy range, $\sim 4\times 10^{-5}$ for the $\tau$ lepton and much smaller for electron and $\mu$. From Eq.~\eqref{eq:21} we also see that at the one loop level the lepton masses depend only on electromagnetic interactions through the coupling $g_{1}$.

The relations $K_{1}$ and $K_{4}$ plotted with the two loop solutions
are no more constant: see Figs.~\ref{fig:1} and~\ref{fig:3}b. For this reason we extended our analysis to the two loop renormalization group equations. At this level the equations become more complicated and it is not possible to \textit{analytically derive} exact evolution constants. However, using the one loop equations we were able to find several relations, which have very small variation during the evolution. For quarks, the evolution of these constants, $K_{2}$ and $K_{3}$, is shown in Fig.~\ref{fig:2} and the variation of $K_{2}$ is of the order of~4\% and that of $K_{3}$ is of the order of~1\%. In the case of two loops the product the quark masses depends on three gauge couplings.

We were also able to obtain a two loop expression for leptons. This generalized expression $K_{5}$ describes the evolution of the charged lepton masses as functions of $g_{1}$ and~$g_{2}$. The variation of $K_{5}$ is of the order of 0.1\% and is very similar for all leptons (see~Fig.~\ref{fig:4}).

Summarizing, we have found several relations between the quark and lepton masses and gauge couplings, which remain (almost) constant upon the renormalization group evolution. Remarkably, all our relations are between the quark and lepton masses and the gauge couplings, but do not contain explicitly the Higgs quartic coupling $\lambda$ or the Higgs field vacuum expectation value $v$. It is also interesting to note that our relations contain the product of the two flavor invariants: determinants of the quark Yukawa coupling matrices~\cite{ref:19,ref:20}. Our analysis clarifies the picture of the renormalization group flow of the Standard Model, which is governed by a set of coupled non-linear differential equations.

\section*{Acknowledgments}
Supported in part by \textit{Proyecto SIP:20161034 y SIP:20170819, Secretaría de Investigación y Posgrado, Beca EDI y Comisión de Operación y Fomento de Actividades Académicas (COFAA) del Instituto Politécnico Nacional (IPN), Mexico}. P.K would also like to thank Professor Duane Dicus for kind hospitality at the Department of Physics, University of Texas at Austin, where part of the work on the paper has been done.

\appendix
\section{Differential equation for determinant}\label{app:A}

Let us suppose that $A$ and $T$ are square matrices of the same dimension and~$A$ fulfills the following differential equation
\begin{equation}\label{eq:a1}
\dfrac{dA}{dt}=A\cdot T.
\end{equation}
We will derive from Eq.~\eqref{eq:a1} the differential equation for $\det A$, the determinant of the matrix $A$.

The Jacobi formula for derivative of a determinant reads
\begin{equation}\label{eq:a2}
\dfrac{d\det A}{dt} =\operatorname{Tr}\left(\operatorname{adj}(A) \dfrac{dA}{dt}\right),
\end{equation}
where $\operatorname{adj}(A)$ is \textit{adjugate} of the matrix $A$ with the following property
\begin{equation}\label{eq:a3}
A\cdot\operatorname{adj}(A)=\operatorname{adj}(A)\cdot A= \det A\cdot I,
\end{equation}
and $I$ is the identity matrix. If we insert Eq.~\eqref{eq:a1} into Eq.~\eqref{eq:a2} and use Eq.~\eqref{eq:a3} then we immediately obtain the differential equation for the determinant of the matrix $A$
\begin{equation}\label{eq:a4}
\dfrac{d\det A}{dt}=\operatorname{Tr}(T)\det A.
\end{equation}

Suppose now that the matrix $A$ is not hermitian and let us consider the hermitian matrix $H=A^{\dagger}A$. From Eq.~\eqref{eq:a1} it is easy to show that matrix $H$ fulfills the following differential equation
\begin{equation}\label{eq:a5}
\dfrac{d H}{dt}=H\cdot T + T^{\dagger}\cdot H.
\end{equation}
Using again the Jacobi formula~\eqref{eq:a2} for the derivative of the determinant we obtain the differential equation for~$\det H$
\begin{equation}\label{eq:a6}
\dfrac{d\det H}{dt}=\operatorname{Tr}(T+T^{\dagger})\det H.
\end{equation}

\end{document}